\documentclass[manuscript,screen]{acmart}

\usepackage{multirow}
\usepackage[linesnumbered,ruled,vlined]{algorithm2e}
\usepackage{algorithmic}
\usepackage{amsmath}
\usepackage{enumitem}
\usepackage{booktabs}
\usepackage{makecell}
\usepackage[flushleft]{threeparttable}

\newcommand{\vpara}[1]{\vspace{0.2cm}\noindent\textbf{#1 }}

\AtBeginDocument{%
  \providecommand\BibTeX{{%
    \normalfont B\kern-0.5em{\scshape i\kern-0.25em b}\kern-0.8em\TeX}}}

\setcopyright{acmlicensed}
\copyrightyear{20XX}
\acmYear{20XX}
\acmDOI{XXXXXXX.XXXXXXX}

\acmJournal{TOIS}
\acmVolume{0}
\acmNumber{0}
\acmArticle{000}
\acmMonth{0}

\acmISBN{978-1-4503-XXXX-X/XX/XX}

\begin{document}

\title{RevGNN: Negative Sampling Enhanced Contrastive Graph Learning for Academic Reviewer Recommendation}

\author{Weibin Liao}
\email{mrblankness@gmail.com}
\orcid{0000-0002-9682-9934}
\authornote{W. Liao and Y. Zhu contributed eaqually to this work.}
\affiliation{%
  \institution{School of Computer Science and Technology, Beijing Institute of Technology}
  \city{Beijing}
  \country{China}
  \postcode{100081}
}

\author{Yifan Zhu}
\orcid{0000-0002-7695-1633}
\authornote{Correspondance to Y. Zhu and X. Li.}
\affiliation{%
  \institution{School of Computer Science, Beijing University of Posts and Telecommunications}
  \city{Beijing}
  \country{China}
  \postcode{100876}
}
\email{yifan_zhu@bupt.edu.cn}

\author{Yanyan Li}
\orcid{0000-0003-3886-1749}
\affiliation{%
  \institution{School of Computer Science and Technology, Beijing Institute of Technology}
  \city{Beijing}
  \country{China}
  \postcode{100081}
}

\author{Qi Zhang}
\orcid{0000-0002-1037-1361}
\email{zhangqi_cs@tongji.edu.cn}
\affiliation{%
 \institution{Department of Computer Science, Tongji University}
 \city{Shanghai}
 \country{China}
 \postcode{200092}
}

\author{Zhonghong Ou}
\orcid{0000-0001-9987-3175}
\affiliation{%
  \institution{State Key Laboratory of Networking and Switching Technology, Beijing University of Posts and Telecommunications}
  \city{Beijing}
  \country{China}
  \postcode{100876}
}
\email{zhonghong.ou@bupt.edu.cn}

\author{Xuesong Li}
\orcid{0000-0003-1570-277X}
\affiliation{%
  \institution{School of Computer Science and Technology, Beijing Institute of Technology}
  \city{Beijing}
  \country{China}
  \postcode{100081}
}
\email{lixuesong@bit.edu.cn}

\renewcommand{\shortauthors}{Liao et al.}

\begin{abstract}
Acquiring reviewers for academic submissions is a challenging recommendation scenario.
Recent graph learning-driven models have made remarkable progress in the field of recommendation, but their performance in the academic reviewer recommendation task may suffer from a significant false negative issue. This arises from the assumption that unobserved edges represent negative samples.
In fact, the mechanism of anonymous review results in inadequate exposure of interactions between reviewers and submissions, leading to a higher number of unobserved interactions compared to those caused by reviewers declining to participate.
Therefore, investigating how to better comprehend the negative labeling of unobserved interactions in academic reviewer recommendations is a significant challenge.
This study aims to tackle the ambiguous nature of unobserved interactions in academic reviewer recommendations. Specifically, we propose an unsupervised Pseudo Neg-Label strategy to enhance graph contrastive learning (GCL) for recommending reviewers for academic submissions, which we call RevGNN. RevGNN utilizes a two-stage encoder structure that encodes both scientific knowledge and behavior using Pseudo Neg-Label to approximate review preference. Extensive experiments on three real-world datasets demonstrate that RevGNN outperforms all baselines across four metrics. Additionally, detailed further analyses confirm the effectiveness of each component in RevGNN.
\end{abstract}



\received{XX XXX XXX}

\maketitle

\section{Introduction}
Academic reviewer recommendation is the process of identifying suitable reviewers for various submissions, such as funding proposals, research papers, technical plans, etc. 
The recommendation model for this task is designed to solve the serious problem of information overload faced by the review system \cite{DBLP:conf/kdd/MimnoM07, DBLP:journals/cacm/PriceF17, DBLP:journals/tetc/CaglieroGPB21}. 
When dealing with numerous submissions and a lengthy list of candidate reviewers, an efficient and accurate method is required to identify experts who possess the requisite skills and are willing to participate in the review process.

Previous reviewer recommendation studies have typically been based on the assumption that reviewers are more likely to review submissions that closely align with their research focus \cite{DBLP:journals/ijitdm/WangSC10}. 
However, this assumption may not always hold due to the latent reviewing preference  \cite{DBLP:journals/asc/YangLYCN20,DBLP:journals/cas/HoangNCH21}.
For example, a reviewer's decision to review a submission is influenced not only by its direct relevance to their area of expertise but also by factors such as the reputation of the venue, the implied relevance of the field, and the reviewer's current workload. Therefore, capturing comprehensive contextual information requires a perspective that considers the interconnectivity among heterogeneous entities.
Therefore, learning better review preferences from reviewers' past behaviors (i.e., reviewer-submission interaction graph) can significantly supplement the shortcomings brought by content-based recommendations.
By leveraging powerful message-passing mechanisms to model users and items with their preference into low-dimensional feature representations within an end-to-end framework, the graph neural networks (GNN)-based approach has been applied to various recommendation tasks, including online shopping, e-learning, and entertainment \cite{pinsage, NGCF, DBLP:journals/kbs/ZhuLLSQN21, zhu2021recommending}, and also has shown significant potential in academic reviewer recommendation.

\begin{figure}[!t]
	\centering
		\includegraphics[width=0.75\textwidth]{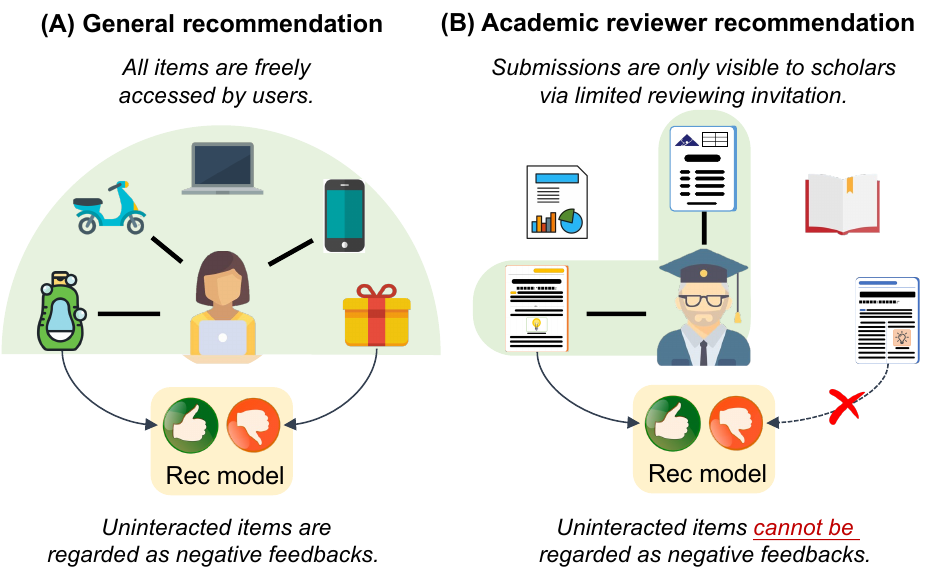}
	\caption{Different visibility of items between (A) general and (B) academic reviewer recommendation.}
	\label{fig:example}
\end{figure}

\vpara{Key Challenge:}
However, previous GNN models cannot be directly used in the academic reviewer recommendation due to their inherent graph sparseness.
As shown in Figure \ref{fig:example}, the majority of peer review records are kept confidential to ensure the objectivity of the review process. This policy makes information sharing among editorial offices and funding agencies nearly impossible \cite{DBLP:journals/cii/LiuWMS16}. Consequently, it is arbitrary to assume that an unobserved interaction between a submission and a reviewer represents a negative sample indicating that the reviewer would not review the submission.
Therefore, the challenge namely \textbf{"Vague meaning of negative interactions"} lies in finding a way to integrate the dense knowledge information in submissions with the limited sparse behavior labels among reviewers \cite{DBLP:journals/corr/abs-2203-15876, DBLP:conf/kdd/YuY000H21}.

In light of the above discussion, this study proposes RevGNN, a novel approach for academic reviewer recommendation using graph contrastive learning (GCL). RevGNN employs a two-stage encoder structure to learn comprehensive embedding representations of reviewers and submissions. In the first stage, we use a decoupled GNN to encode the behavior preference of reviewers, while the prior scientific semantic knowledge of submissions is captured using a pre-trained scientific language model. In the second stage, we address the extreme sparsity in the scholar-submission bipartite graph by introducing the Pseudo Neg-Label strategy to enhance the performance of negative sampling during the graph contrastive learning process.
The major contributions of this study are summarized as follows:
\begin{itemize}
[leftmargin=*,itemsep=0pt,parsep=0.5em,topsep=0.3em,partopsep=0.3em]
    \item Facing the observation that recommendations based solely on topic similarity were far inefficient, we propose a graph learning-based recommendation framework namely RevGNN which comprehensively considers the reviewer's behavioral preferences, submission's implicit prior knowledge, and their contrastive features.
    \item We argue that the role of unobserved interactions is not the same as general recommendation due to insufficient exposure, thereby causing false negative sampling. Thus we propose the Pseudo Neg-Label strategy to extract appropriate negative samples during the GCL training.
    \item We perform extensive experiments on three public real-world datasets, and the results show that our proposed RevGNN outperforms all baselines over four metrics.
\end{itemize}
 
\section{Related Works}

\subsection{Academic Reviewer Recommendation}

As formally introduced by \cite{DBLP:journals/tois/ZhangZDCZT20}. the academic reviewer's recommendation refers to suggesting the editor/administrator with experts in one or several fields to jointly evaluate the scientific research proposals or achievements (such as funding proposal, paper, and technical reports). 
An efficient academic reviewer recommendation is urgently needed under the massive number of manuscripts, as a vivid exampple is that the number of submission received by NeurIPS (Annual Conference on Neural Information Processing Systems) and AAAI (The Association for the Advancement of Artificial Intelligence) in 2020 is 5.6 and 5.5 times respectively than that in 2014 \cite{DBLP:conf/wsdm/Shah21}.
Currently, the academic reviewer recommendation has become the cornerstone of submission management systems among academic journals, conferences and funding bodies, to reduce the unbearable workload of finding suitable reviewers who also agree to provide the review service \cite{DBLP:journals/ipm/ZhaoZ22}.
There are usually two types of methodology for academic reviewer recommendation \cite{DBLP:journals/ijitdm/WangSC10}, which are retrieval-based and assignment-based approaches.

The retrieval-based approach focuses on improving the matching degree between the submission and reviewer candidates. 
This type of recommendation emphasizes the significance of relevance and coverage, and takes a loose basic hypothesis of the application scenario:  Reviewers are free to choose whether or not to review the submission.
Early retrieval-based approaches are mostly based on content matching with probabilistic topic models.
\cite{DBLP:conf/sigir/BalogAR06} proposed two Bayesian probability formulas to calculate the expertise of an expert on a given document.
\cite{DBLP:conf/sigir/FangSM10} calculated the conditional probability of relevance given a pair of queries and scholar, and the calculation considered heterogeneous features.
The language models are utilized to model the content, such as LDA \cite{DBLP:journals/kbs/DaudLZM10} and LSA \cite{DBLP:journals/kbs/ProtasiewiczPKD16}.
Recently, relevance matching has extended to semantic matching.
For example, 
\cite{DBLP:journals/ir/TanDZCZ21} proposed a semantic-based iterative model with deep representation learning which enables direct learn the embeddings of the text information, and used deep neural networks to conduct end-to-end reviewer recommendations based on these embeddings.
\cite{DBLP:journals/tois/ZhangZDCZT20} transferred the academic reviewer recommendation as a multi-label classification task, and proposed a bidirectional Gate Recurrent Unit to predict the relationship between submissions and reviewers with the intermediary discipline concepts.

Assignment-based approaches transfer the academic reviewer recommendation into an assignment-based optimization problem. 
It takes a more strict assumption that reviewers are obliged to review the assigned submissions.
In this setting, a recommendation is made by following soft and hard constraints defined by the model design, and bad cases are tolerant for the purpose of the global optimum.
There are several ways to construct constraints, such as market bidding strategies \cite{DBLP:conf/aaai/MeirLLMK21}, random constraints \cite{DBLP:conf/nips/JecmenZLSCF20}, and local fairness \cite{DBLP:conf/kdd/KobrenSM19}. 
However, both approaches ignore the comprehensive profiling modeling, including review behavior and relative differences between reviewers and submissions.
These implicit factors are also as critical to review as content relevance and assignment efficiency.

\subsection{Graph Learning-based Recommendation}

The graph neural network, as a new type of neural network, has been proposed to extract features from non-Euclidean space data. GNNs capture the complex relationships between nodes and edges in graph-structured data, making them applicable to various scenarios such as knowledge graphs~\cite{mu2021knowledge,ma2023kr}, recommendation systems~\cite{LightGCN,SVDpp,ApeGNN,zhang2024recdcl}, and time series prediction~\cite{yi2024fouriergnn}. The work most relevant to ours involves user recommendation using GNNs, particularly those studies based on bipartite graphs of historical interactions between users and items. These studies employ graph-based learning methods to learn embedding representations and predict user preferences for unseen items\cite{GNNRec_survey}.
Early matrix factorization methods, such as SVD++ \cite{SVDpp}, decompose the adjacency matrix of the graph to derive the embedding, which is useful but insufficient for encoding complex user behaviors.
Then, the idea of neural collaborative filtering, such as NeuMF \cite{NeuMF}, CML \cite{CML} and NIRec \cite{NIRec}, is proposed to replace the inner product of matrix factorization with parameter-learnable neural networks.
Emerging Graph Convolution Networks (GCN) are also used to learn embeddings via the feature propagation of graph convolution, and achieve convincing performance in both theoretical and practical research, such as PinSage \cite{pinsage}, NGCF \cite{NGCF}, and LightGCN \cite{LightGCN}.

Recently, the self-supervised contrastive learning studies have been reported to alleviate the sparse cold-start interactions in graph recommendation \cite{DBLP:journals/corr/abs-2203-15876}, and received massive attention.
SGL \cite{SGL} augmented node representations by dropping out elements on the manipulated graphs.
SEPT \cite{DBLP:conf/kdd/YuY000H21} proposed a social recommender system that generates multiple positive samples for contrastive learning on the perturbed graph. 
SimGCL simplifies the contrastive learning paradigm in the recommendation by replacing the graph augmentations with uniform noises for contrastive views \cite{DBLP:conf/sigir/YuY00CN22}.
However, it should be noted that graph learning-based recommendation models have not been widely adopted into academic reviewer recommendations yet, due to the performance degradation caused by the false negative sampling on sparse graphs with low exposure rates.

\section{Preliminaries}

\subsection{Problem Formulation}

In this study, we define the academic reviewer recommendation as a link prediction task in a heterogeneous graph.
That is, we define a heterogeneous academic graph $\mathcal{G} = \{ \mathcal{N}, \mathcal{E} \}$ where the $\mathcal{N} = \{ N_{scholar}, N_{submission}, ... \}$ denotes the set of different types of nodes, and $\mathcal{E}=\{ E_{review}, E_{author}, ...\} $ denotes the set of different types of relationships. Let the number of nodes $|\mathcal{N}| = n$, and we have $ E_i \subset \mathcal{N} \times \mathcal{N}$ where $i$ denotes any type of possible relations. The objective of academic reviewer recommendation is to construct a function $f$ that can predict if there exists a ``review" relation by giving a submission-scholar pair:

\begin{equation}
    \begin{aligned}
        & Find \quad f(u,v) \longrightarrow \{0, 1\}, \\
        & s.t. \quad f(u,v) = 1 \Leftrightarrow e_{u,v} \in E_{review}, \\ 
        & where \quad u \in N_{scholar}, \ v \in N_{submission}
    \end{aligned}
    \label{eq:task_formulation}
\end{equation}

It should be noted that the submission and scholar nodes, alongside with review relationship between these two types of nodes are necessary for our task, and other types of nodes, as well as relations, are optional but useful in portraiting submissions and scholars in the body of knowledge.

Table~\ref{tab:symbols} lists the notations of symbols and their descriptions that are used throughout this paper.

\begin{table}[!t]
    \centering
    \caption{Notations of Symbols and Their Descriptions Used in This Paper}
    \begin{tabular}{l l} 
      \hline
      Notations & Descriptions \\
      \hline
      $\mathcal{G}$ & The heterogeneous academic graph \\
      $\mathcal{N}$ & The set of different types of nodes in $\mathcal{G}$ \\
      $\mathcal{E}$ & The set of different types of relationships in $\mathcal{G}$ \\
      $\mathcal{U} \in \mathbb{R}^m$ & The set of users in GNN-based recommendation model, $m$ denotes the number of users \\
      $\mathcal{V} \in \mathbb{R}^n$ & The set of items in GNN-based recommendation model, $m$ denotes the number of items  \\
      $\mathcal{R} \in \mathbb{R}^{m \times n}$ & The set of behavioral interactions in GNN-based recommendation model \\
      $u$ & A user instance from $\mathcal{U}$ \\
      $v$ & A item instance from $\mathcal{V}$ \\
      $\textbf{h}_u^{(l)}$ & The representation of node $u$ in the $l$-th layer of the GNN \\
      $\sigma(\cdot)$ & The non-linear activation function \\
      $\textbf{W}$ &  Weight matrix for model training \\
      $\tilde{\textbf{A}} = \textbf{D}^{-\frac{1}{2}}\textbf{A}\textbf{D}^{-\frac{1}{2}}$ & \makecell[l]{The normalized adjacent matrix, $\textbf{A}$ is the adjacent matrix with self-loops and $\textbf{D}$ is the \\ corresponding diagonal degree matrix}\\
      $L$ & The total number of propagation layers \\
      $d$ & The  number of embedding dimensions \\
      $y_{u,v}$, $\hat{y}_{u,v}$ & The ground truth / predict probability of edge from submission $u$ to scholar $v$ \\
      $C$ & The number of clusters in clustering \\
      $\{\mu_i\}$ & The embedding in the clustering of the $i$-th cluster \\
      \hline 
    \end{tabular}
    \label{tab:symbols}
\end{table}

\subsection{GNN-based Recommendation}
The GNN-based recommendation model usually works on a user-item bipartite graph, denoted as $\mathcal{G}=\{\mathcal{U}, \mathcal{V}, \mathcal{R} \}$ which consists of users $\mathcal{U} = \{u_1, u_2, \dots, u_m\}$, items $\mathcal{V} = \{v_1, v_2, \dots, v_n\}$, behavioral interactions $\mathcal{R} \in \mathbb{R}^{m \times n}$, where $m = |\mathcal{U}|$ and  $n=|\mathcal{V}|$.
The objective of the GNN-based model is to predict whether there is an edge by giving a pair of specific users and items. 
Taking a similar foundation with other GNN models, GNN-based recommendation models conduct message passing over the bipartite graph to derive the semantic representations of users and items. 
Generally, this message passing has two processes, namely the aggregation and pooling \cite{GNNRec_survey}. 

In the aggregation process, the GNN model collects representation propagated by the neighbor for each node $u$ at each layer, and this process at $l$-layer can be represented as:
\begin{equation}
    \textbf{h}_u^{(l)} = U(\textbf{h}_u^{(l-1)}, AGG(\{\textbf{h}_v^{(l-1)}, \forall {v} \in \mathcal{N}_u)\}),
    \label{eq:update}
\end{equation}
\noindent where $AGG$ and $U$ stand for the specific aggregation and update strategies.
For instance, GCN~\cite{GCN} aggregates each node's neighborhood nodes by using the following propagation rule:

\begin{equation}
\textbf{h}_u^{(l)} = \sigma(\textbf{W}^{(l)}(\textbf{h}_u^{(l-1)} + \tilde{\textbf{A}}\textbf{h}_v^{(l-1)})) 
\end{equation}

\noindent where $\textbf{h}_u^{(l)}$ denotes the embedding of node $u$ after $l$-layer's aggregation, $\sigma(\cdot)$ denotes the non-linear activation function, $\textbf{W}$ denotes a weight matrix, and $\mathcal{N}_u$ denotes the neighborhood of node $u$.
$\tilde{\textbf{A}} = \textbf{D}^{-\frac{1}{2}}\textbf{A}\textbf{D}^{-\frac{1}{2}}$, where $\textbf{D}$ is the diagonal node degree matrix with self loop $\textbf{D}_{ii}=\sum\nolimits_{j}A_{ij}+ \textbf{I}$,  
note that $\textbf{I}$ is the identity matrix. 

Then, during the Pooling operation, the representation learned by the GNN is utilized to perform prediction. 
Usually, these pooling functions combine the embeddings propagated at each layer to produce the final representation $\textbf{h}_u^{*}$. 
Formally, for node $u$, the the pooling function $POOL$ can be formulated as:
\begin{equation}
    \textbf{h}_u^{*} = POOL(\textbf{h}_u^{(0)}, \textbf{h}_u^{(1)}, \cdots, \textbf{h}_u^{(L)}),
    \label{eq:pooling}
\end{equation}
\noindent where $L$ is the total number of propagation layers.

\subsection{Graph Contrastive Learning for Recommendation}

The basic idea of Contrastive Learning is to construct a particular set of representational invariances as data augmentations. 
Moreover, GCL takes an additional hypothesis that changing embedding representation and partial structure perturbation are independent \cite{DBLP:conf/sigir/YuY00CN22}. 
Thus, in addition to the original supervised recommendation loss, the GCL-based recommendation model introduces a supervised contrastive loss.
Taking SGL \cite{SGL} as example, it utilizes an InfoNCE \cite{DBLP:journals/corr/abs-1807-03748} loss:

\begin{equation}
    \mathcal{L}_{CL}= \sum_{i \in \mathcal{B}} = - \log \frac{e^{z_{i}^{'\top}z_{i}^{''}/\tau}}
    {\sum_{j \in \mathcal{B}} e^{z_{i}^{'\top}z_{j}^{''}/\tau}},
    \label{eq:InfoNCE}
\end{equation}
where $z_{i}$, $z_{j}$ are users/items in the same sampled batch $\mathcal{B}$, $z^{'}$ and $z^{''}$ are augmented node representations learned from two dropout downsampled graphs, and $\tau > 0$ is the temperature. 
The CL loss tries to optimize the representation of nodes, making the distance of similar positive pairs ($z_{i}^{'}$ and $z_{i}^{''}$) closer, while the dissimilar negative pairs ($z_{i}^{'}$ and $z_{j}^{''}$) more orthogonal.

\section{The Proposed RevGNN}

\subsection{Overview}

\begin{figure*}[!t]
	\centering
		\includegraphics[width=0.99\textwidth]{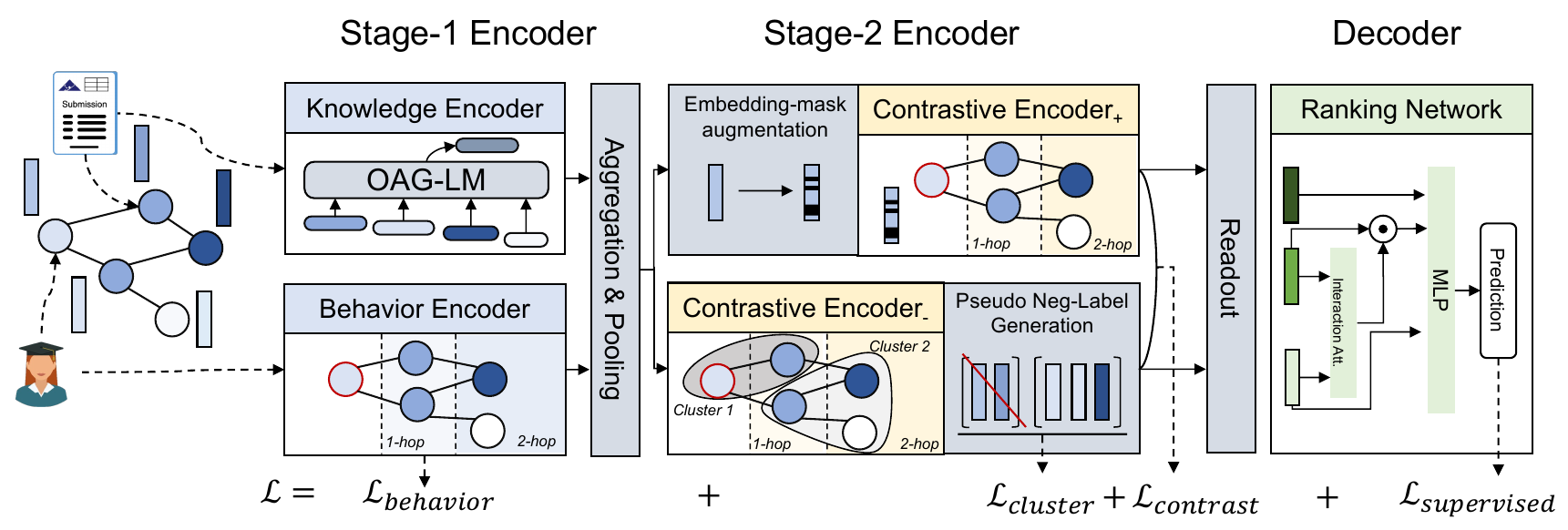}
	\caption{The framework of RevGNN.}
	\label{fig:framework}
\end{figure*}

We present the general framework of the proposed RevGNN in Figure \ref{fig:framework}. 
RevGNN has a two-stage encoder and one recommendation decoder. 
In the encoding process, the inputted heterogeneous graph is first detached into a bipartite to represent the submission-scholar behavior and a knowledge graph to describe the knowledge of nodes.
We derive the basic embedding representation of nodes with two structure-specific encoders, namely preference encoder and knowledge encoder, according to the graphical type, and fuse them again with the aggregation operation. 
Then, in the second stage, we propose to perform graph contrastive learning by manipulating positive embedding and extracting negative samples with clustering pseudo labels in two contrastive encoders.
Based on the readout operation, we utilize an interaction-based neural network structure to predict the edge existence based on the embedding of the scholar-submission pair.

\subsection{Stage-1 Encoder: Behavior \& Knowledge Representation Learning}

The heterogeneous academic graph provides not only informative features for describing entities including submissions and scholars but also difficulty in modeling the multi-modal structure. 
Thus, we propose a ``divide and rule"-based strategy to model different types of components in this graph.

The RevGNN first detaches the entire academic graph into two subgraphs: 
\begin{itemize}
[leftmargin=*,itemsep=0pt,parsep=0.5em,topsep=0.3em,partopsep=0.3em]
    \item \textbf{Submission-scholar bipartite graph:} This subgraph only contains scholar, submission, and review relations between them. Considering that review behavior is an action-based behavior, this bipartite graph provides an indication of the preferences and habits that scholars (i.e., potential reviewers) have maintained over the past review process.
    \item \textbf{Academic knowledge graph:} This subgraph equals the complementary set of the submission-scholar bipartite graph. Most relations and attributes in this graph are fact-based behaviors, therefore it is considered to provide subject information and relevance in the current knowledge system.
\end{itemize}
Then, these two subgraphs are encoded by two completely different encoders to extract corresponding features.

\vpara{Preference Encoder.} 
For the submission-scholar bipartite graph, we utilize a decoupled graph convolution network (GCN) \cite{SGC} to effectively learn the representation embedding of submission and scholar by aggregating embedding from their neighbor nodes.  
Formally, we denote this encoding process via a $l$-layer decoupled GCN as:
\begin{equation}
\boldsymbol{h}_{u}^{b,(l)} =  \hat{\boldsymbol{A}} \boldsymbol{h}^{b,(l-1)} \boldsymbol{W}_b^{(l)}. 
\label{eq:hl}
\end{equation}
where $\boldsymbol{H}_b^{(l)} = [\boldsymbol{h}_{1}^{b,(l)}, ..., \boldsymbol{h}_{n}^{b,(l)}]^{T} \in \mathbb{R}^{n \times d}$ is the representation of the nodes derived at $l$-th layer, $\hat{\boldsymbol{A}} = \boldsymbol{D}^{-1/2} \boldsymbol{A} \boldsymbol{D}^{-1/2}$ denotes the normalized adjacent matrix given that $\boldsymbol{A}$ is the adjacent matrix with self-loops and $\boldsymbol{D}$ is the corresponding diagonal degree matrix.
$\boldsymbol{W}_b^{(l)}$ is a learnable parameter matrix at $l$-th layer and $d$ denotes the number of embedding dimensions.
The reason for choosing this decoupled graph convolution is to avoid using the activation function, so as to enable the entire training process into one matrix multiplication process with better efficiency \cite{DecoupledGCN, DecoupledGCN-ECCV}.
The probability of edge from submission $u$ to scholar $v$ is predicted through a non-parametric residual dot production:
\begin{equation}
\hat{y}^b_{u,v} = ReLU(\boldsymbol{h}^b_{u} \odot \boldsymbol{h}^b_{v}),
\label{eq:dot_production}
\end{equation}
where
\begin{equation}
        \boldsymbol{h}^b_{u} = \sum^{l}_{i=1} \boldsymbol{h}_{u}^{b,(i)}, \quad
        \boldsymbol{h}^b_{v} = \sum^{l}_{i=1} \boldsymbol{h}_{v}^{b,(i)}.
\label{eq:residual}
\end{equation}

Unlike GNN-based recommendation models such as LightGCN \cite{LightGCN} and NGCF \cite{NGCF}, we do not use the Bayesian Personalized Ranking loss (BPR) \cite{BPR} as the optimization loss function, and the reasons are twofold. 
First, the basic assumption of BPR is that the unobserved edge in a graph is very likely to be a negative sample, which does not meet the fact of the sparsity issue in the submission-scholar bipartite graph. 
Second, the sampling time brought by BPR is likely to be redundant, which is to be addressed in the Stage-2 Encoder of RevGNN. 
Thus, we select the binary cross-entropy loss (BCE) for the preference encoder on observed edges only:
\begin{equation}
\mathcal{L}_{beh}=-\sum_{(u,v) \in E_{review}} {y}_{u,v} \log \hat{y}_{u,v}^b + (1-{y}_{u,v}) \log (1 - \hat{y}_{u,v}^b),
\label{eq:bce_loss}
\end{equation}
where ${y}_{u,v}$ denotes the observed edges between $u$ and $v$. 

\vpara{Knowledge Encoder.}
The text title and abstract of a submission provide a refined theme description with scientific key points. 
However, the understanding of this theme, such as the location in the disciplinary taxonomy, exists in an implicit rather than an explicit form. 
Therefore, it is necessary to consider not only the title and abstract but also to combine the prior knowledge to learn the relevance of different kinds of relations for comprehensive representation learning.

Here we utilize the OAG-LM (also known as OAG-BERT) \cite{OAG-BERT}, a scientific domain-specific language encoding model, to produce the embedding representation for each submission. 
Different from the traditional plain text scientific pre-training model (Such as SciBERT \cite{SciBERT}), OAG-LM designs a two-dimensional location coding, so that we can input heterogeneous entity-relationship type information into the additional dimension, and the embedding of this information could also be inputted into the OAG-LM as prior knowledge. 
According to its design, in addition to the title, heterogeneous entities and relations including abstract, venue, and authoring scholar affiliations are able to be optional graphic knowledge to improve the representation performance.
In other words, the full formulation of $\mathcal{N}$ and $\mathcal{E}$ in the problem formulation are $\mathcal{N} = \{ N_{scholar}, N_{submission}, N_{venue}, N_{affiliation}\}$ and $\mathcal{E} = \{ E_{review}, E_{author}, E_{affilied}, E_{published}\}$.
The output of OAG-LM is the embedding of the [CLS] token, which is regarded as storing the information of all inputted data.
However, as a pre-trained model, we do not finetune the OAG-LM with submission record in the experiment to avoid data leakage problems, and thus the loss would not be used for optimization during the training process. 

\vpara{Aggregation and Pooling.}
After knowledge and behavioral information are encoded separately, we use specific strategies to aggregate them together as the basis for the next stage of encoding.
Also, for efficiency reasons, we need to keep only those elements that are directly related to the review purpose.
Therefore, the encoding objective of this stage is to transfer heterogeneous node and relationship information to scholars and submissions, and we compress the representation of scholars and submissions into a fixed low-dimensional space, which is denoted as:
\begin{equation}
        \boldsymbol{h}^1_{u} = AGG(\boldsymbol{h}^b_{u}, \boldsymbol{h}^{k}_{u}),
\label{eq:aggregation}
\end{equation}
where $\boldsymbol{h}^k_{u}$ is the embedding representation produced by the knowledge encoder, and $AGG$ is the aggregation operation. 
This study uses concatenation for simplification, i.e., $\boldsymbol{h}^1_{u} = \boldsymbol{h}^b_{u} || \boldsymbol{h}^{k}_{u}$. 
In particular, when node $u$ is a scholar, its corresponding knowledge embedding is derived from the average pooling of all of its immediate submission neighbors because the theme of scholars is reflected by the items they interacted.

To summarize, in this Stage-1 Encoder, we optimize $\mathcal{L}_{beh}$ to tune the parameters in decoupled GCN layer $W_b$, and we encode the interaction as well as prior knowledge into node embedding representations with the help of pre-trained scientific language model.

\begin{figure}[!t]
	\centering
		\includegraphics[width=0.90\textwidth]{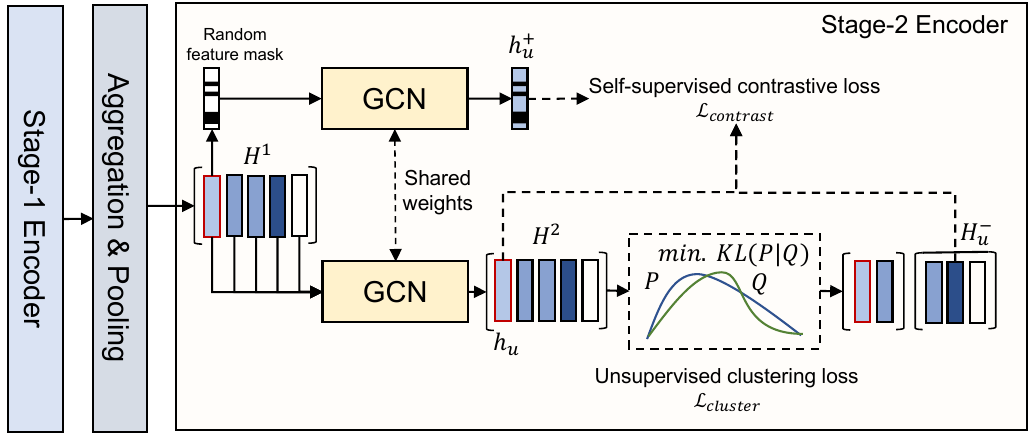}
	\caption{The stage-2 encoding process. In this stage, the embedding representation of nodes is adjusted by graph contrastive learning. Particularly, we design a Pseudo Neg-Label Sampling strategy to avoid false negative edges for the observation-insufficient scholar-submission interaction graph.}
	\label{fig:stage-2}
\end{figure}

\subsection{Stage-2 Encoder: Contrastive Representation Learning with Pseudo Neg-Label}

In this section, we propose a contrastive encoding stage for better relative embedding representation learning under a graph that is extremely sparse but has lots of unobserved false negative edges.
We first construct the positive sample strategy by manipulating the embedding representation of the current node's neighbors. 
Furthermore, inspired by the idea of Debiased Contrastive Learning \cite{GDCL}, we propose an unsupervised predecessor task during the negative sampling, namely Pseudo Neg-Label, to assign nodes into different classes, and sample negative nodes with different classes for the current node. 

\vpara{Rethinking the Sampling in GCL Strategy.}
The key challenge of academic reviewer recommendation is that the scholar-submission interaction graph is extremely sparse due to the review mechanism, and this sparsity is likely to come from unobserved edges. 
Although the recent development of self-supervised learning methods on graphs makes node representations more effective by performing relative embedding learning by constructing self-generated labels on graphs, most of them still assume that an unobserved edge tends to be a non-existent edge (i.e., a negative sample) \cite{liu2022graph,xie2022self}. 
In fact, in our task, the reason why the reviewer node did not have a review edge with the submission node may be that the reviewer had not been exposed to the submission before. 
The performance of the existing comparative learning method would degrade our task. 
The negative sampling strategy becomes the bottleneck of the current graph contrastive learning task on this type of graph \cite{DBLP:conf/kdd/YangDZYZT20}.
Therefore, how to select appropriate negative samples for scholars and submissions in this contrastive encoding stage, thereby making the self-supervised task both correct and difficult, is the main challenge to be addressed.

\vpara{Contrastive Encoder with Pseudo Neg-Label Sampling.}
The graph contrastive learning's key idea is to perform relative representation learning on pairs of data points.
It tries to optimize the representation of nodes, making the distance of similar (positive) pairs closer, while the dissimilar (negative) pairs more orthogonal.
Therefore, the strategy of extracting positive and negative samples for the current node to establish similar and negative pairs is the key design in graph contrastive learning.

Most of the existing recommendation models based on graph contrastive learning focus on how to construct positive samples.
For the selection of negative samples, the existing methods often focus on how to improve the difficulty of distinguishing negative samples from positive samples, so that the model can learn more features from difficult negative samples \cite{yang2022region}.
However, in our academic reviewer recommendation task, we need to focus on avoiding potentially false negative samples. 
Therefore, we propose to construct mediation pseudo-tags according to data distribution and select only nodes inconsistent with the current node pseudo-tags as negative samples.
The full pipeline of this process is shown in Figure \ref{fig:stage-2}.

Formally, in RevGNN, we use a common approach to derive a positive sample by manipulating feature embedding on each node and using a GCN layer \cite{GCN} to propagate embedding through the graph:
\begin{equation}
        \boldsymbol{h}^{+}_{u} = \sigma ( \boldsymbol{\hat{A}}  \boldsymbol{h}^{1}_{u'} \boldsymbol{W}_{c} ) ,
\label{eq:positive_sample}
\end{equation}
where $\boldsymbol{h}^{1}_{u'}$ is derived by a function that randomly masks a portion of embedding on the given node \cite{GRAND}, $\boldsymbol{W}_{c}$ is the weighted matrix, and $\sigma(\cdot)$ is the activation function. 

On the other hand, to avoid false negative sampling issues, we design a clustering layer on the embedded representation of nodes to jointly optimize with the graph contrastive learning framework. 
We first use the shared-weight GCN layer as used in positive sampling, which is denoted as:
\begin{equation}
        \boldsymbol{H^2} = \sigma ( \boldsymbol{\hat{A}}  \boldsymbol{H^1} \boldsymbol{W}_{c} ) ,
\label{eq:GCN_nodrop}
\end{equation}
where $H^1$ is the matrix of all nodes' embedding ($h_u^1 \in H^1$).
The clustering layer provides each node with a cluster ID which is able to be regarded as a discriminative self-supervised pseudo label. 
Suppose there are $C$ clusters, and their cluster center $\{\mu_i\}_{i=1}^{C}$ are randomly initialized in the space of $\mathrm{R}^{d}$, the clustering is performed by minimzing the following KL-divergence-based cluster loss:
\begin{equation}
    \mathcal{L}_{clus} = \sum_{u} \sum_{i=1}^{C} p_{u,i} \log \frac{p_{u,i}}{q_{u,i}},
\label{eq:cluster_loss}
\end{equation}
where $i$ denotes the $i$-th cluster, and $u \in N_{scholar} \cup N_{submission}$. 
$p_{u, i}$ is the target distribution which is calculated by:
\begin{equation}
    p_{u,i} = \frac{q_{u,i}^2 / \sum_{u} q_{u,i}}{ \sum_{i} q_{u,i}^2 / \sum_{u} q_{u,i} },
\label{eq:pui}
\end{equation}
where $q_{u,i}$ denotes the similarity between embedding representation $h_u$ and cluster center $\mu_i$, which is measured by the $t$-distribution: 
\begin{equation}
    q_{u,i} = \frac{\sum_i (1+\Vert h_u - \mu_i \Vert^2)}{1+\Vert h_u - \mu_i \Vert^2}.
\label{eq:qui}
\end{equation}
By minimizing $\mathcal{L}_{clus}$, we can assign each node with a cluster ID. 
We regard this ID as a mediation pseudo-tag, and therefore in the negative sampling we only select nodes from clusters in which the current node is not located, thereby avoiding adding potential false negative nodes.
We name this process the Pseudo Neg-Label sampling, and here we have the final negative sample set:
\begin{equation}
    H_{u}^{-} = \{h_v\}_{c(u) \neq c(v)},
\label{eq:h-}
\end{equation}
where $c(u)$ denotes the cluster ID of node $u$.

Based on the derived positive sample $h_u^+$ (Eq. \ref{eq:positive_sample}), negative sample set $ H_{u}^{-}$ (Eq. \ref{eq:h-}), and the original encoded embedding $h_u$, we construct the InfoNCE-like contrastive loss:
\begin{equation}
    \mathcal{L}_{cl} = \sum_{h_u \in H^2} -\log \frac{e^{h_u^T h_u^+}}{e^{h_u^T h_u^+} + \sum_{h_v \in H_u^-}  e^{h_u^T h_v}}.
\label{eq:contrastive_loss}
\end{equation}

To summarize, in this stage-2 encoder, we jointly optimize $\mathcal{L}_{cl}$ and $ \mathcal{L}_{clus}$ to tune the parameters in GCN layer $W_c$, and thereby encoding the contrastive mutual information into node embedding representations.

\subsection{Interaction-Oriented Recommendation Decoder}

After the two-stage encoding, scholars and submissions are adequately represented by embedding vectors in a fixed dimension.
Then, a decoder is required to resolve the embedding representation to predict whether there is an edge by giving any pair of scholars and submissions.
Considering that embedding representations contain behavior, prior knowledge, and comparative information, we design a ranking-based recommendation network that can calculate the interactive attention between submissions that the current and reviewers have interacted with in the past. 

\begin{figure}[!t]
	\centering
	\includegraphics[width=0.4\textwidth]{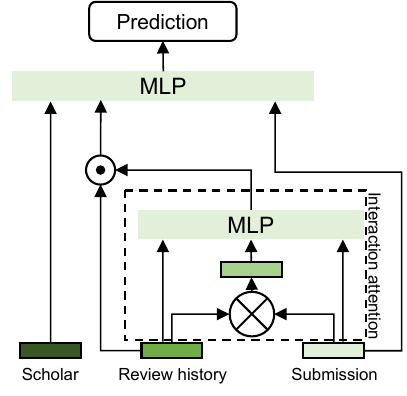}
	\caption{The network structure of the decoder. The decoder network predicts the probability of review action by considering the attention between the candidate submission and the scholar's historical review.}
	\label{fig:DIN}
\end{figure}

As shown in Figure \ref{fig:DIN}, the interaction attention is actually a variant of the activation unit in the design of Deep Interest Network (DIN) \cite{DIN}.
In the beginning, we provide the encoded embedding representations of a scholar (potential reviewer), submissions the scholar has reviewed, and the submission candidate. 
We denote them as $h_u$, $H_p = h_p^1, h_p^2,...$, and $h_v$, respectively.
Note that the history of review could have multiple submissions, as we only draw the scenario where there is only one submission for the purpose of concision.

\begin{algorithm}[!t]
\small
\SetKwInOut{Input}{Input}\SetKwInOut{Output}{Output}
	\Input{Heterogeneous academic graph $\mathcal{G}$, Node feature $X$} 
	\Output{Optimized model parameters $\mathbf{\Theta}$}
	\BlankLine 
	 \emph{Initialize $\mathbf{\Theta}$} \; 
      \While{$\mathbf{\Theta}$ does not converge}{
                \tcp{\textcolor[rgb]{0.5,0.6,0.7}{Stage-1 Preference Encoder}} 
                Calculate the behaviour loss: $\mathcal{L}_{beh} \leftarrow$ Eq. \eqref{eq:bce_loss} \;
                Derive the preference embedding: $\boldsymbol{h}_{u}^{b} \leftarrow$ Eq. \eqref{eq:hl} \;
                \tcp{\textcolor[rgb]{0.5,0.6,0.7}{Stage-1 Knowledge Encoder}} 
                Derive the knowledge embedding: $\boldsymbol{h}_{u}^{k} \leftarrow \text{OAG-LM}(u,v)$ \;
                \tcp{\textcolor[rgb]{0.5,0.6,0.7}{Aggregation and Pooling}} 
                $\boldsymbol{h}^1_{u} = AGG(\boldsymbol{h}^b_{u}, \boldsymbol{h}^{k}_{u})$\;
                \tcp{\textcolor[rgb]{0.5,0.6,0.7}{Stage-2 Encoder}} 
	        \For{$i: 1 \to m $}{
                    \tcp{\textcolor[rgb]{0.5,0.6,0.7}{Positive Sample}} 
                    $\boldsymbol{h}^{+}_{u} \leftarrow$ Eq. \eqref{eq:positive_sample} \;
                    cluster \;
                    \For{$i: 1 \to C $}{
                        \tcp{\textcolor[rgb]{0.5,0.6,0.7}{Calculate the cluster loss}} 
                        Calculate the cluster loss: $\mathcal{L}_{clus} \leftarrow$ Eq. \eqref{eq:cluster_loss} \;
                        \tcp{\textcolor[rgb]{0.5,0.6,0.7}{Pseudo Neg-Label Sample}} 
                        $\boldsymbol{h}^{-}_{u} \leftarrow$ Eq. \eqref{eq:h-}  \;
                        \tcp{\textcolor[rgb]{0.5,0.6,0.7}{Calculate the contrastive loss}} 
                        Calculate the cluster loss: $\mathcal{L}_{cl} \leftarrow$ Eq. \eqref{eq:contrastive_loss}  \;
                    }
                }
                \tcp{\textcolor[rgb]{0.5,0.6,0.7}{Interaction-Oriented Recommendation Decoder}} 
                Calculate the recommendation loss: $\mathcal{L}_{sup} \leftarrow$ Eq. \eqref{eq:loss_decoder} \;
                Update RevGNN: $\mathbf{\Theta} = \mathcal{L} + \nabla \mathcal{L} \leftarrow$ Eq. \eqref{eq:loss} \;
	 }
	 Return $\mathbf{\Theta}$.
      \caption{RevGNN training algorithm}
      \label{RevGNN_Algorithm_training} 
\end{algorithm}

In the decoding stage, we first calculate the interaction attention between the candidate submission and the historical submissions.
For each historical submission, the out product (Kronecker product) is performed, and then we concatenate it with the original $h_p^i$ and $h_v$ to a multilayer perceptron layer (MLP):
\begin{equation}
    a_i = W_a^{(2)} \sigma ( W_a^{(1)} (h_p^i \otimes h_v ) + b_a^{(1)} ) + b_a^{(2)},
\label{eq:out_product}
\end{equation}
where $b_a^{(1)}$ and $b_a^{(2)}$ are bias matrix, $W_a^{(1)}$ and $W_a^{(2)}$ are weighted matrix with the size of $(d^2, \eta)$ and $(\eta, 1)$, and $\eta$ is the size of hidden dimension. 
We also select the PReLU \cite{DBLP:conf/iccv/HeZRS15} as the activation by following the design of DIN. 
The output $a_i$ is regarded as an attentive weight that scales the impact of historical review. 

In the final recommendation, we use another MLP to predict the probability between scholar and submission with the concatenation of scholar, submission, and weighted historical submissions:

\begin{equation}
        \hat{y}_{u,v} = Sigmoid(\sigma( W_r^{(2)} \sigma ( W_r^{(1)} (h_u || \sum_i a_i h_p^i || h_v) \\ + b_r^{(1)} ) + b_r^{(2)})).
\label{eq:final_prediction}
\end{equation}

The similar BCE loss is selected as the negative log-likelihood loss function to back-propagate gradients to the weighted matrices in the decoder:
\begin{equation}
    \mathcal{L}_{sup} = -\sum_{(u,v) \in E_{review}} {y}_{u,v} \log \hat{y}_{u,v} + (1-{y}_{u,v}) \log (1 - \hat{y}_{u,v}).
\label{eq:loss_decoder}
\end{equation}

Finally, the general loss of RevGNN is formulated as:
\begin{equation}
    \mathcal{L} = \mathcal{L}_{beh} +  \mathcal{L}_{clus} +  \mathcal{L}_{cl} + \mathcal{L}_{sup},
\label{eq:loss}
\end{equation}

Eventually, he entire training process of RevGNN is summarized in Algorithm \ref{RevGNN_Algorithm_training}.

\section{Experiments}
\subsection{Datasets}

Due to the confidentiality of peer review, publicly available benchmarks are extremely scarce. Previous studies have mostly evaluated using simulated data, such as constructing datasets through expert self- or third-party annotation. 
Although this approach is reasonable for retrieval tasks, it may disrupt the density and other characteristics of the actual review behavior of reviewer recommendation scenarios. Therefore, we opted for the recently released Frontiers dataset \cite{OAG-Bench}, sourced from real peer review, as the primary dataset to validate our proposed RevGNN model. 
This dataset is a recently released public repository for evaluating the performance of finding reviewers for paper submissions \footnote{\url{https://github.com/THUDM/Reviewer-Rec}}.

It collects paper information from six Frontiers Press journals with high-impact factors, as well as reviewer information disclosed by their editorial office.
Due to a lack of complete information and limited hardware capacity, we reshuffled the entire data records to produce smaller but high-quality subsets. 
We retain high-quality reviewers with at least 2 or 3 reviews and formed Frontiers-4k and Frontiers-8k.
Frontiers-4k has 4,000 submissions and 6,711 reviewers, and each submission has at least two reviewers. 
Similarly, Frontiers-8k has 8,132 submissions and 9,560 reviewers where each submission has at least two reviews.

To demonstrate the generalizability of our proposed RevGNN across different scenarios, we also test the traditional NIPS dataset to validate the model's performance. The NIPS dataset is an earlier released, small-scale, manually annotated dataset \cite{NIPSDataset}. The NIPS dataset \cite{NIPSDataset} consists of rankings of reviewers and accepted papers from the NIPS 2006 conference, obtained through expert human judgments of relevance. These rankings are used to approximate the assignment of reviewers to the submitted articles. The dataset is widely utilized in research related to reviewer recommendations for peer review of research articles \cite{DBLP:journals/ir/TanDZCZ21}.

\begin{table}[!t]
\caption{Statistics of Frontiers-4k, Frontiers-8k and NIPS.}
\label{tab:statistics}
\centering
\begin{tabular}{p{3.3cm}p{1.6cm}p{1.6cm}p{1.6cm}}
\toprule
 Item & \multicolumn{1}{r}{Frontiers-4k} & \multicolumn{1}{r}{Frontiers-8k} & \multicolumn{1}{r}{NIPS}\\
\midrule
Release Year & \multicolumn{1}{r}{2023} & \multicolumn{1}{r}{2023} & \multicolumn{1}{r}{2008} \\
 \# Scholar & \multicolumn{1}{r}{6,711} & \multicolumn{1}{r}{9,560} & \multicolumn{1}{r}{364}\\
 \# Submission & \multicolumn{1}{r}{4,000} & \multicolumn{1}{r}{8,132} & \multicolumn{1}{r}{13,138}\\
 \# Review & \multicolumn{1}{r}{10,799} & \multicolumn{1}{r}{19,063} & \multicolumn{1}{r}{29,731} \\ 
 Density & \multicolumn{1}{r}{$4.02\times 10^{-4}$} & \multicolumn{1}{r}{$2.45 \times 10^{-4}$} & \multicolumn{1}{r}{$6.22 \times 10^{-3}$} \\ 
 \# Affiliation \& venue & \multicolumn{1}{r}{8,272} & \multicolumn{1}{r}{11,872} & \multicolumn{1}{r}{N/A} \\ 
 \# Other edges & \multicolumn{1}{r}{12,267} & \multicolumn{1}{r}{19,999} & \multicolumn{1}{r}{N/A} \\
\bottomrule
\end{tabular}
\end{table}

When testing the recommendation performance, for each submission, we randomly select one of the reviewers to the test set and keep the rest as the training set. 
The statistics of the three datasets are shown in TABLE \ref{tab:statistics}.
As a comparison, The commonly used benchmark dataset in general recommendation studies, MovieLens-1m, has 6,040 users and 3,883 movies, and there are 1,000,209 edges between users and movies, with a density of $4.26 \times 10^{-2}$.
The difference in density also indicates the extreme sparsity in the academic reviewer recommendation, and this sparsity is the inevitable result of the review mechanism, thereby confirming our hypothesis about the false negative label of the unobserved edge.

\subsection{Baselines}

To verify the superiority of the proposed RevGNN model in academic reviewer recommendation, we compared the performance of the following baseline models:

\begin{itemize}
    \item TF-IDF Reviewer Rec \cite{DBLP:journals/kbs/ProtasiewiczPKD16}: TF-IDF is a classic statistical method used to assess the importance of a word to a set of documents. It is introduced into the expert matching problem by measuring the consine distance on TF-IDF between the reviewer and the submitting paper.
    \item TPMS~\cite{Toronto}: Toronto Paper Matching System (TPMS) algorithm is a tool commonly used for academic conference paper assignment. It calculates matching scores by analyzing the research interests and domain keywords of authors and reviewers, thereby optimizing the match between papers and reviewers to improve the quality and efficiency of the review process.
    \item Wide\&Deep \cite{WandD}: Wide \& Deep is another classic click-through rate (CTR)-based recommendation model which is composed of a Wide part of a single layer and a Deep part of a multi-layer. It enables a strong memory ability and generalization ability, and finally can quickly process and remember a large number of historical behavior features in the process of recommendation ranking.
    \item DeepFM \cite{DeepFM}: DeepFM is an improvement based on Wide\&Deep. By combining the Deep part with the Factorization Machine (FM), the model replaces the Wide part with the FM layer, which improves the model's ability to extract information from the embedded representation. 
    \item ENMF \cite{ENMF}: ENMF is a matrix factorization-based recommendation model with a non-sampling strategy. It argues that a negative sampling strategy is not always robust, thus the performance of sampling-based methods could degrade in some cases.
    \item RecVAE \cite{RecVAE}: RecVAE is a variational autoencoder-based recommendation model by including composite prior distribution between the encoding and decoding. 
    \item LightGCN \cite{LightGCN}: LightGCN is a representative GNN-based recommendation model that simplifies the graph convolution network by removing modules to establish matrix-based embedding propagation and update.
    \item SGL \cite{SGL}: SGL is a state-of-the-art self-supervised graph learning recommendation model. It introduces an auxiliary self-supervised task to pre-train the node representation via self-discrimination and then applies to the semi-supervised specific task.
    \item SimGCL \cite{SimGCL}: SimGCL is a CL-based model that learns unified user/item representations for recommendation via the contrastive learning paradigm, thereby implicitly alleviating popularity bias. Additionally, SimGCL discards the graph augmentation mechanism and instead adds uniform noise to the embedding space to create contrastive views.
    \item SHT \cite{SHT}: SHT is another CL-based model that employs a transformer based on hypergraphs to achieve global collaborative filtering for users and items. Additionally, it introduces a cross-view generative self-supervised learning component for data augmentation of the interaction graph between users and items.
    \item ApeGNN \cite{ApeGNN}: ApeGNN is a diffusion-based GNN approach that designs a node-wise adaptive mechanism for message aggregation so that each node can adaptively decide its weights based on the local degree structure. In the experiment, we select both local structure settings and denote them as HK (Heat kernel) as well as PPR (Personalized PageRank).

\end{itemize}

\subsection{Evaluation Metrics}
In this study, we select \textit{Recall}, \textit{Precision}, \textit{Hit Rate}, and \textit{Normalized Discounted Cumulative Gain} to measure the performance of academic reviewer recommendation.
\begin{itemize}
[leftmargin=*,itemsep=0pt,parsep=0.5em,topsep=0.3em,partopsep=0.3em]
    \item \textbf{Recall} evaluates the coverage of effective items in the prediction list of top-K recommendations.
    \begin{equation}
        Recall@K = \frac{1}{|\mathcal{N}_{submission}|}\sum\limits_{{u_i}~\in~\mathcal{N}_{submission}}\frac{|T({u_i})|~\cap~|R({u_i})|}{|T({u_i})|}, s.t.|R({u_i})|=K,
    \end{equation}
    \noindent where $T({u_i})$ and $R({u_i})$ are the ground and predicted reviewer set of submussion $u_i$ respectively.
    \item \textbf{Precision} evaluates the availability of effective items in the prediction list of top-K recommendations.
    \begin{equation}
        Precision@K = \frac{1}{|\mathcal{N}_{submission}|}\sum\limits_{{u_i}~\in~\mathcal{N}_{submission}}\frac{|T({u_i})|~\cap~|R({u_i})|}{|R({u_i})|}, s.t.|R({u_i})|=K.
    \end{equation}
    \item \textbf{Normalized Discounted Cumulative Gain (NDCG)} measures the quality of higher-ranked true positives. It is calculated based on the discounted cumulative gain (DCG) which is derived by:
    \begin{equation}
     DCG@K =\frac{1}{|\mathcal{N}_{submission}|}\sum\limits_{{u_i}~\in~\mathcal{N}_{submission}}\sum\limits_{k=1}^{K}\frac{2^{(rel_{k,{u_i}})} -1}{log_2(2+k)},
     \end{equation}
    \noindent where $rel_{k,{u_i}}$ is 1 if $k$-th recommended reviewer is the ground truth reviewer for the submission ${u_i}$, otherwise it is 0. Then the NDCG@K is computed by:
    \begin{equation}
     NDCG@K = \frac{DCG@K}{IDCG@K},
     \label{eq:NDCG}
    \end{equation}
    \noindent where IDCG@K is the ideal cumulative gain.

    \item \textbf{Hit Ratio (HR)} evaluates the proportion of the total number of ground truths in the actual top-K recommendation list.
    \begin{equation}
    Hit~Ratio@K=\frac{\sum\limits_{i~\in~\mathcal{N}_{submission}}R(i)}{\sum\limits_{i~\in~\mathcal{N}_{submission}}T(i)}.
    \end{equation}
\end{itemize}

Specifically, we evaluate the results on the top-20 recommendation list and denote these metrics as R@20, P@20, N@20 and HR@20.

\subsection{Experiment Setup}

To achieve the comparison of our RevGNN approach with the aforementioned baselines, we adopted the following experimental setup scheme.
First, we implement all baseline models by using the unified recommendation library called RecBole \cite{recbole}.
We conducted all experiments on a Ubuntu 18.04.2 LTS server with AMD EPYC 7642 48-Core Processor, shared 1008 GB RAM, and an NVIDIA A100 Tensor Core GPU. 
The RevGNN is implemented with PyTorch 1.9.1, and Python 3.9.7.
In the experiment, we select the ReLU function as the activation for Eq. \ref{eq:positive_sample}, \ref{eq:GCN_nodrop}, and set $\eta = 36$ for Eq. \ref{eq:out_product}. The size of $W_r^{(2)}$ and $W_r^{(1)}$ in Eq. \ref{eq:final_prediction} are $(3d, 200)$ and $(200, 80)$, respectively, where $d=932$.
Moreover, for the Stage-1 Encoder, we utilize a 3-layer GCN to perform encoding, while for the Stage-2 Encoder, a 1-layer GCN is employed. We use the Adam optimizer with varying learning rates to optimize the different structures of RevGNN. Specifically, we set the learning rate to 0.001 for both Stage-1 Encoder and Recommendation Decoder, and 0.0001 for Stage-2 Encoder.

We also conduct the Whitney-Mann U-test between the results of multiple runs of RevGNN and the best baseline methods to statistically test the significance of performance improvements by using different random seeds. The significant results with $p$-value $<$ 0.001 are marked with $\dagger$ in the tables.

\subsection{Main Results}

\paragraph{RevGNN on Frontiers-4k \& Frontiers-8k dataset}

We compare the performance of academic reviewer recommendations for paper submissions on Frontiers-4k and Frontiers-8k datasets between RevGNN and all baselines. 
The main result is shown in TABLE \ref{tab:main_result}. 
For each metric, we highlight the best results with \textbf{bold} and denote the second best result with \underline{underline}. 
In addition, the "N/A" mark in the table indicates the corresponding number is less than 0.0001.

\begin{table}[!t]
    \centering
    \caption{Overall performance comparison between RevGNN and baselines. Marker $\dagger$ indicates the statistical significance between RevGNN and the most competitive baseline ($p$-value $<$ 0.001).}
    \label{tab:main_result}
    \begin{tabular}{c|c c c c|c c c c} 
      \hline
      Dataset & \multicolumn{4}{c|}{Frontiers-4k} &\multicolumn{4}{c}{Frontiers-8k}  \\
      \hline
      Metrics & R@20 & N@20 & HR@20 & P@20 & R@20 & N@20 & HR@20 & P@20 \\
      \hline
      TF-IDF & 0.0122 & 0.0084 & N/A & 0.0022 & N/A & N/A & N/A & N/A \\
      TPMS &  0.0802 &  0.0264 & 0.0832 & 0.0043 &  0.0546 & 0.0187 & 0.0573 & 0.0029 \\
      Wide\&Deep & 0.0132 & 0.0054 & 0.0267 & 0.0014 & 0.0070 & 0.0036 & 0.0140 & 0.0007\\
      DeepFM & 0.0118 & 0.0045 & 0.0238 & 0.0013 & 0.0070 & 0.0034 & 0.0140 & 0.0007\\
      ENMF & 0.0123 & 0.0074 & 0.0253 & 0.0013 & 0.0465 & 0.0224 & 0.0963 & 0.0049\\
      RecVAE & \underline{0.0814} & 0.0384 & \underline{0.1664} & \underline{0.0086} & \underline{0.0789} & \underline{0.0418} & \underline{0.1584} & \underline{0.0083}\\
      LightGCN & 0.0457 & 0.0219 & 0.0936 & 0.0048 & 0.0422 & 0.0213 & 0.0841 & 0.0044\\
      SGL & 0.0524 & 0.0257 & 0.1070 & 0.0053 & 0.0409 & 0.0202 & 0.0835 & 0.0043\\
      SimGCL & 0.0470	& 0.0212 & 0.0949 & 0.0049 & 0.0383 & 0.0184 & 0.0772 & 0.0039 \\
      SHT & 0.0623 & 0.0364 & 0.1262 & 0.0053 & 0.0557 & 0.0337 & 0.1123 & 0.0030 \\
      ApeGNN\_HK & 0.0676 & \underline{0.0420} & 0.0684 & 0.0035 & 0.0572 & 0.0322 & 0.0603 & 0.0030 \\
      ApeGNN\_PPR & 0.0705 & 0.0416 & 0.0713 & 0.0036 & 0.0477 & 0.0287 & 0.0512 & 0.0026 \\
      \hline
      RevGNN & \textbf{0.1076}$^{\dagger}$ & \textbf{0.0522}$^{\dagger}$ & \textbf{0.2140}$^{\dagger}$ & \textbf{0.0116}$^{\dagger}$ & \textbf{0.0967}$^{\dagger}$ & \textbf{0.0470}$^{\dagger}$ & \textbf{0.1895}$^{\dagger}$ & \textbf{0.0100}$^{\dagger}$\\
      \%Improv. & 32.19\% & 24.29\% & 28.61\% & 34.88\% & 22.56\% & 12.44\% & 19.63\% & 20.48\% \\
      \hline
    \end{tabular}
\end{table}

Based on the results, it is observed that the proposed RevGNN achieves optimal performance in all four metrics and on both datasets. 
In particular, we analyze our results with each type of recommendation approach:
Traditional topic-matching approaches (such as TF-IDF) have been proved useful in many paper assignment tasks but perform extremely poorly in both datasets. However, this type of approach is feasible only if a limited pool of submissions and reviewers is given, and the reviewers review the assigned papers unconditionally. Therefore, in the absence of mandatory constraints, we take the reviewer's actual behavior feedback as the label to find that the relationship between the reviewer's decision and the subject matching is not causal. In RevGNN, we utilize the semantic knowledge from the pre-trained scientific model, which establishes a more comprehensive subject encoding on both textual and meta-knowledge of the submission.
In addition, the TPMS algorithm demonstrates a disappointing performance. We attribute this to its limited approach of extracting keywords from the vocabulary and constructing embeddings, coupled with inadequate similarity calculation methods. The embedding calculation for keywords lacks contextualization, resulting in significant information loss. Furthermore, TPMS relies solely on the user's historical review queue for modeling, disregarding the user's implicit preferences. In contrast, RevGNN achieves more effective review recommendations by dual modeling of user behavior and knowledge, thereby uncovering deep representations of user profiles. This improvement is evidenced by a 34.16\% increase in R@20.

The CTR-based recommendation methods (Wide\&Deep and DeepFM) also have below-average performance. Because the CTR model requires a large amount of data training to complete the learning of the embedding, insufficient data volume without prior knowledge in the reviewer recommendation scenario leads to serious overfitting issues. In RevGNN, we also use a ranking-based decoder, but before that, we have to process the embedding representation of reviewers and submissions via a two-stage encoder to fully learn the behavior, knowledge, and contrastive relation.

The non-sampling-based MF method (ENMF) also has degraded performance, and the reason is similar to the CTR-based approach. Based on this result, we show again that many hypotheses including unobserved edges may not be applicable to the academic reviewer recommendation scenario. In RevGNN, we show that the negative sampling strategy can provide sufficient performance improvement in the case of a missing hypothesis for the unobserved edge.

The autoencoder-based approach (RecVAE) surprisingly has a very competitive performance. The idea of the autoencoder is to manipulate or compress the initialized embedded representation through the encoder, and then try to reconstruct through the decoder. In addition to the supervised label, the model also benefits from external label-free learning to enhance the generalization ability. Similarly, the RevGNN also emphasizes that appropriate and effective self-supervised tasks should be designed in the case of sparse labels to enhance the generalization of encoding and improve the R@20 by 32.19\% compared to RecVAE.

For graph learning-based approaches (LightGCN), we observe a significant above-average performance on both datasets. This result shows that the reviewer's behavioral preference is an important factor in impression review behavior, and it also shows that this behavioral preference can be learned by graph learning. However, due to the extreme absence of available labels, the observable edges that graph learning relies on are not enough for representation propagation and aggregation, which leads to model performance deficiency. RevGNN also applies GNN structure to the encoder in the two stages of behavior and contrastive learning, and further introduces self-supervised representation learning, thereby reducing the dependence on external labels.

Self-supervised approach (SGL, SimGCL, and SHT) takes advantage of the generalization ability when the external label is sparse in a graph. 
SGL enhances performance based on a pre-training strategy. However, SGL only considers the self-supervision task of positive samples, neglecting the contribution of negative samples, and therefore degrading in our cases. 
SimGCL adds uniform noise to the embedding space to create contrastive views, thereby maximizing the consistency of node representations across different graph augmentations, without considering the differential representation of negative samples. 
Similarly, SHT explores global collaborative relationships explicitly for data augmentation on the user-item interaction graph. 
In contrast, RevGNN designs a negative sampling-enhanced GCL strategy to improve the representation capability of GCN on negative samples with varying contributions, resulting in a 105.34\% improvement in R@20 compared to SGL.

Graph diffusion-based ApeGNN also achieves a competitive performance in both Frontiers-4k and 8k datasets.
ApeGNN adaptively decides the weights of message aggregation in a node-wise manner, and this process is determined by the local structure (Heat Kernel and Personalized PageRank). 
It indicates that profiling local structure helps GNN to overcome the sparseness without adding learnable parameters. 
However, RevGNN still outperforms ApeGNN by 52.62\% on Frointiters-4k and 102.7\% on Frontiers-8k in terms of R@20, showing the significance of contrastive learning edges are as the label and are extremely sparse.

\paragraph{RevGNN on NIPS dataset}

\begin{table*}[!t]
    \centering
    \caption{Overall performance comparison between RevGNN and baselines. Marker $\dagger$ indicates the statistical significance between RevGNN and the most competitive baseline ($p$-value $<$ 0.001).}
    \label{tab:main_result_nips}
    \begin{tabular}{c|c c c c} 
      \hline
      Dataset & \multicolumn{4}{c}{NIPS} \\
      \hline
      Metrics & R@20 & N@20 & HR@20 & P@20 \\
      \hline
      TF-IDF & N/A & N/A & N/A & N/A \\
      TPMS & 0.0912 & 0.0583 & 0.2104 & 0.0169 \\
      Wide\&Deep & 0.1219 & 0.0610 & 0.2541 & 0.0203 \\
      DeepFM & 0.1219 & 0.0610 & 0.2541 & 0.0203 \\
      ENMF & 0.0070 & 0.0029 & 0.0247 & 0.0012 \\
      RecVAE & 0.1045 & 0.0805 & \underline{0.4890} & \underline{0.0315} \\
      LightGCN & 0.0945 & 0.0543 & 0.2226 & 0.0182 \\
      SGL & 0.0995 & 0.0666 & 0.2253 & 0.0125 \\
      SimGCL & \underline{0.1408} & \underline{0.0880} & 0.4411 & 0.0242 \\
      SHT & 0.1142 & 0.0564 & 0.2396 & 0.0110 \\
      ApeGNN\_HK & 0.1267 & 0.0721 & 0.2962 & 0.0179 \\
      ApeGNN\_PPR & 0.1024 & 0.0754 & 0.2469 & 0.0154 \\
      \hline
      RevGNN & \textbf{0.1526}$^{\dagger}$ & \textbf{0.1246}$^{\dagger}$ & \textbf{0.6099}$^{\dagger}$ & \textbf{0.0420}$^{\dagger}$ \\
      \%Improv. & 8.38\% & 41.59\% & 24.72\% & 33.33\% \\
      \hline
    \end{tabular}
\end{table*}

We further compare the performance of academic reviewer recommendations for paper submissions on the NIPS dataset between RevGNN and all baselines. 
The main result is shown in TABLE \ref{tab:main_result_nips}. 
Similar to the results on Frontiers, we highlight the best results with \textbf{bold}, and denote the second best result with \underline{underline} for each metric.

We observed similar results on the NIPS dataset as we did on the Frontiers dataset, where RevGNN outperformed all baseline models. This was validated across four metrics, further demonstrating that RevGNN benefits from behavior modeling and knowledge modeling for mining data information, along with advanced negative sampling optimization strategies, enhancing the expressive power of GNNs.

Furthermore, we found that RecVAE exhibited competitive performance on the NIPS dataset once again, albeit limited to the HR@20 and P@20 metrics. However, for the R@20 and N@20 metrics, the self-supervised SimGCL achieved performance second only to RevGNN. This once again confirms our conclusion from the Frontiers dataset that in the case of sparse labels, it is crucial to design appropriately effective self-supervised tasks to enhance the generalization capability of encoding.

It should be noted that NIPS is an old dataset as we can even know from its name. 
In this dataset, only 364 candidate reviewers labeled more than 13 thousand submissions, which is severely out of line with reality.
While this labeling is suitable for training retrieval models, it fails to simulate the sparsity of reviewer recommendations in real-world scenarios.
Correspondingly, the Frontiers dataset is directly collected from real-world peer review scenarios, and its sparsity better reflects the actual reviewer recommendations than the NIPS dataset. 
Therefore, in the subsequent in-depth analyses, we mainly report the results from Frontiers as the primary dataset.

\begin{table*}[!t]
  \begin{center}
    \caption{Ablation Result of RevGNN.}
    \label{tab:ablation}
    \scalebox{1.0}{
    \begin{tabular}{c|c c c c|c c c c} 
      \hline
      Dataset & \multicolumn{4}{c|}{Frontiers-4k} &\multicolumn{4}{c}{Frontiers-8k} \\
      \hline
      Metrics & R@20 & N@20 & HR@20 & P@20 & R@20 & N@20 & HR@20 & P@20\\
      \hline
      -Knowl. & 0.0868 & 0.0431 & 0.1798 & 0.0094 & 0.0813 & 0.0401 & 0.1627 & 0.0084 \\
      -Behav. & 0.0887 & 0.0428 & 0.1798 & 0.0094 & 0.0856 & 0.0439 & 0.1718 & 0.0089\\
      -Stage-2 & 0.0992 & 0.0505 & 0.2021 & 0.0103 & 0.0944 & 0.0451 & 0.1883 & 0.0099\\
      \hline
      RevGNN & \textbf{0.1076} & \textbf{0.0522} & \textbf{0.2140} & \textbf{0.0116} & \textbf{0.0967} & \textbf{0.0470} & \textbf{0.1895} & \textbf{0.0100}\\
      \hline
    \end{tabular}
    }
  \end{center}
\end{table*}

\begin{figure}[!t]
\centering
\includegraphics[width=0.8\textwidth]{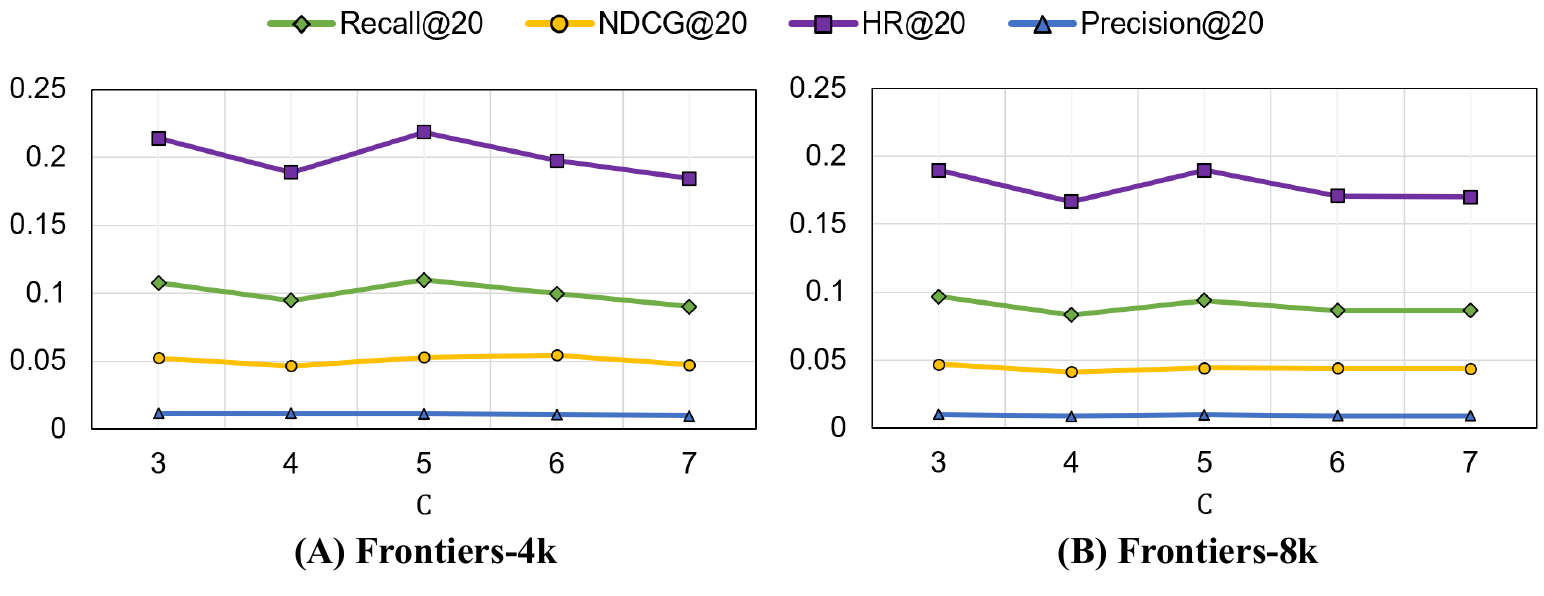}
\caption{The variation of performance with different numbers of cluster $C$.}
\label{fig:number_of_c}
\end{figure}

\subsection{Ablation Study \& Parameter Analysis}

After verifying the effectiveness of RevGNN in academic reviewer recommendations, we further study the contribution of various components of this model. 
Due to the small size of the NIPS dataset, which consists of only 364 scholars, the experimental results lack representativeness. Therefore, we conducted ablation experiments and analysis of RevGNN solely on the Frontiers datasets.
As shown in TABLE \ref{tab:ablation}, we compare the performance variation by removing any component in the RevGNN. 
Here, ``-Knowl." means the RevGNN which removes the knowledge encoder, ``-Behav." indicates the RevGNN which removes the behavior encoder, and ``-Stage-2" refers to the model without the entire stage-2 contrastive learning encoder.
Based on the results, it is observed that the model seems to form a comprehensive representation by utilizing all encoders, meaning that once we remove a certain component the model would suffer a certain degradation of performance.
It is not surprising that the reduction of the knowledge or behavior encoders would lead to a decline in the performance of RevGNN, as this implies a loss of information. Furthermore, we observed that a reduction in the contrastive learning encoder led to a decline in all metrics on both datasets, effectively substantiating the efficacy of our Pseudo Neg-Label strategy.

We also investigate the performance impact caused by the hyper-parameter $C$.
As introduced in the model section, $C$ decides the number of classes the nodes in the scholar-submission bipartite graph would be clustered to.
However, currently, the value of $C$ needs to be manually set, so we further study how the performance will be influenced when changing $C$.
Similar to the analysis of ablation experiments, we performed parameter analysis of RevGNN solely on the Frontiers dataset.
We test different value of $C \in \{3, 4, 5, 6, 7\}$, and present the results in Figure \ref{fig:number_of_c}.
Based on the results, we observe that the performance does not change dramatically with the change in the value of $C$, which indicates the robustness of the RevGNN to this hyperparameter. 
Finally, the model achieves the best performance when $C=5$.
In RevGNN, the purpose of clustering is to find nodes whose distribution is inconsistent with the current node and the specific category of these nodes is not considered. 
So the increase in the number of $C$ does not change the actual negative sample process.

\section{Discussion}

\begin{figure}[!t]
\centering
\includegraphics[width=0.95\textwidth]{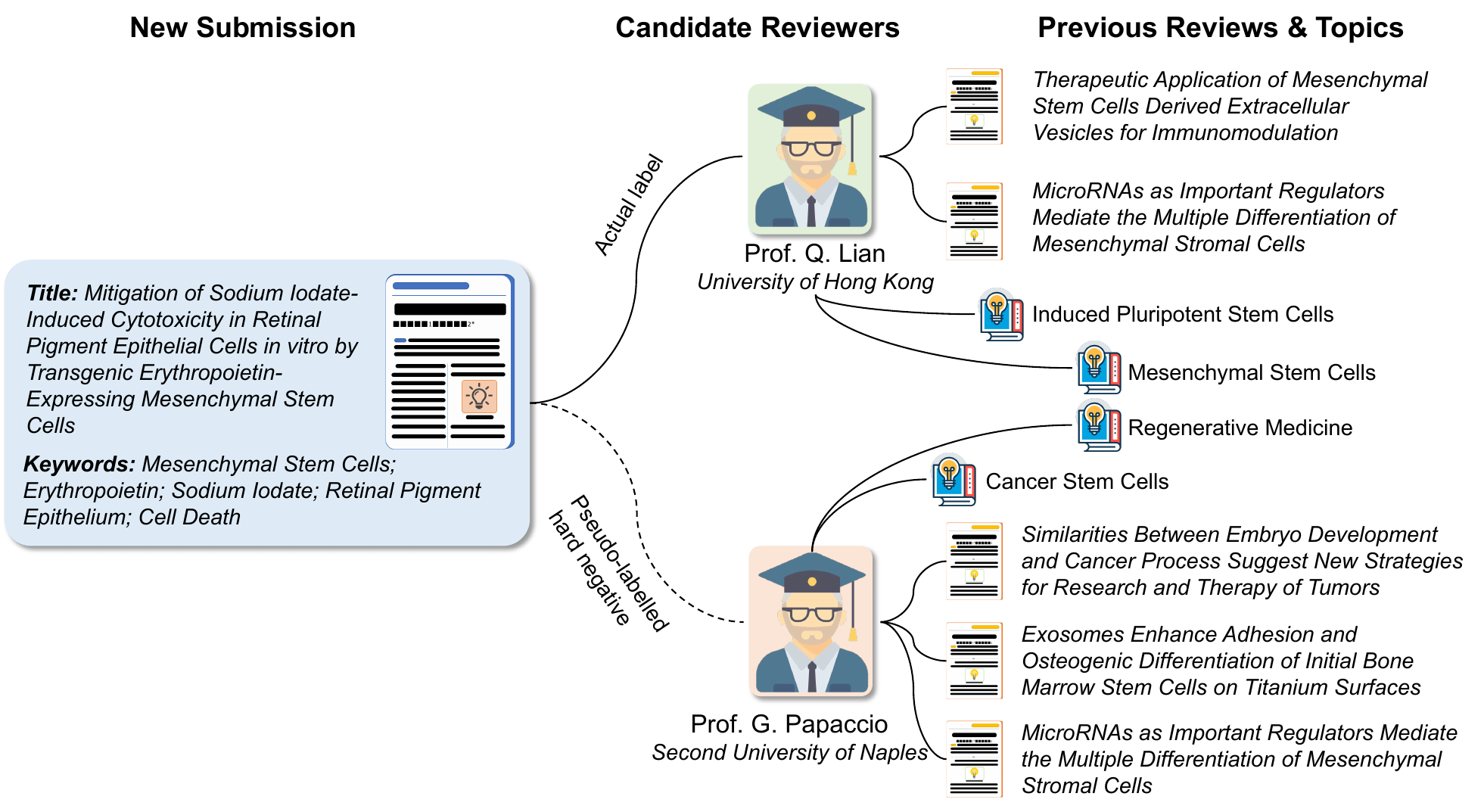}
\caption{Representative case of correctly recommendations by RevGNN on the Frontiers-8k dataset.}
\label{fig:casestudy}
\end{figure}

\subsection{Case Study}

To facilitate better understanding of RevGNN, in Fig.~\ref{fig:casestudy}, we present representative case which correctly recommendations by RevGNN on the Frontiers-8k dataset. Fig.~\ref{fig:casestudy} illustrates a submission titled ``\textit{Mitigation of Sodium Iodate-Induced Cytotoxicity in Retinal Pigment Epithelial Cells in vitro by Transgenic Erythropoietin-Expressing Mesenchymal Stem Cells},'' which describes a specific toxicity during the differentiation of \textit{Mesenchymal Stem Cells} into retinal pigment epithelium leading to cell death. 
Fig.~\ref{fig:casestudy} also depicts two candidate reviewers along with their reviewed submissions, highlighting the relevant topics they have interested. From this figure, it is observed that both candidates are highly reputable senior professors with deep research backgrounds in Mesenchymal and Stem Cells.

For the candidate reviewer \textit{Prof. G. Papaccio}, RevGNN recognizes that although this candidate has research topics similar to the new submission, there are still differences in the distribution of research details compared to the true labels. 
Although the reviewer has research topics related to \textit{Mesenchymal Stem Cells} and \textit{Cancer Stem Cells}. His deeper focus is located in therapy on bone tumour which belongs to clinical usage. 
During the training process, RevGNN considers \textit{Prof. G. Papaccio} as a hard negative sample, and assign a pseudo-negaitve-label to it in the stage 2. 
As a result, RevGNN ultimately recommends \textit{Prof. Q. Lian} as the reviewer for the new submission, an expert more focused on the study of \textit{Mesenchymal Stem Cells} in differentiation and immune mechanisms, achieving accurate recommendation. 
In contrast, all other baseline models failed to make correct recommendations on this submission. 
This case demonstrates the effectiveness of the pseudo-negative label strategy we proposed, enabling more precise judgments among multiple candidate reviewers in similar research fields.

However, it is worth noting that this example is presented, interpreted, and analyzed based on the positive results generated by RevGNN. We must acknowledge the possibility of subjective cognitive biases in these results. We believe that further research and exploration are necessary for interpretable deep recommendation models, especially those based on graph networks.

\subsection{Observations}
Academic reviewer recommendation is a long-term concern of Informetrics and academic data mining research. 
Unlike other review recommendations, such as code review, academic reviewer recommendations need to be able to fully understand the background knowledge of the reviewing items and their complex taxonomy relationships. 
Moreover, the unique anonymous review mechanism of academic review greatly restricts the recommendation model from training to evaluation, making the relevant progress slower than the general recommendation studies.

In this study, we introduce the self-supervised contrastive learning method in the field of graph representation learning to the academic reviewer recommendation problem and propose the RevGNN model to accurately recommend potential reviewers for submissions.
Based on experiments on the three public datasets, our method significantly outperforms all baseline models, thus confirming the effectiveness of the design of RevGNN.
Furthermore, we note that our results are also derived on the basis of the following additional observations and analyses.
\begin{itemize}
    \item Scholars and submissions implicitly contain complex background knowledge. In a relatively open academic field, attempting to calculate the topic similarity of any scholar-submission pair without a systematic knowledge base produces a large number of false negatives. For example, a submission titled ``collaborative filtering of graphs" and an expert on ``web recommendation systems" are not similar in literal terms, but overlap in semantic terms. Fortunately, current semantic modeling methods such as scientific pre-training model \cite{SciBERT,OAG-BERT} and knowledge graph has brought prominent solutions \cite{MAG, DBLP:journals/fgcs/LinZLSN21}. 
    \item The reviewer's behavior in deciding whether to review is not determined solely by the thematic similarity between the reviewer and the submission. In the absence of mandatory constraints and with the discretion of the reviewers to review, we observed that implementing recommendations based solely on topic similarity was far less efficient than what was practical. Therefore, an efficient reviewer recommendation system should also consider reviewers' preferences, including frequency, academic social networks, publications, and institutional reputation \cite{DBLP:journals/cacm/PriceF17}. 
    \item We also argue that the nature of academic reviewer recommendations breaks the basic assumption on unobserved edges in general recommendations. In the general graph recommendation, the fact that the user has an observed interaction with the item is sufficient to indicate the user's preference for it compared to the other items with no observed interactions. However, as mentioned above, the closed and confidential review mechanism does not allow sufficient exposure between scholars and submissions, so the unobserved interaction does not reveal the user's behavioral preferences, and this insufficient exposure also leads to serious graph sparsity. 
\end{itemize}

\subsection{Limitation Study}

Although experimental results show that the proposed RevGNN has achieved a significant performance improvement, it should also be noted that the current study still has the following limitations.

First, the evaluation of our experiment may still suffer from false negative labels. 
Although we adopted the method of Pseudo Neg-Label to alleviate the influence brought by false negative samples in the training process, the evaluation of the experiment can only be measured by the actual review behavior due to the lack of comprehensive and sufficient exposure between reviewers and submissions. 
In other words, the actual recommendation performance of each model should be higher than our reported results. 
It is expensive to have a large number of human experts label submissions for their own review, and there is a serious bias problem with a small number of experts labeling preferences for all scholars.
Therefore, further solutions, such as crowd-sourcing, are needed to obtain widespread and sufficient labels.

Second, the capacity of the model is limited.  
After experimental evaluation, we found that the maximum load capacity of our model on an A100 GPU is comparable to the Frontiers-8k dataset (8,132 submissions and 9,560 scholars), and the GPU memory for storing embedding representations is around 60GB. 
Although this capacity is equivalent to the common GCL models and can cover most academic reviewer recommendation scenarios (such as journals and most academic conferences), it still faces a GPU memory shortage for large-scale recommendations. 
For example, large conferences such as IJCAI (International Joint Conference on Artificial Intelligence) and AAAI (The Association for the Advancement of Artificial Intelligence) attract more than 10,000 submissions, and a similar situation occurs when proposals are reviewed by the National Natural Science Foundation of China in every recent year \cite{DBLP:journals/tmis/SilvaJC14}.
Therefore, our method still needs to try to reduce its dependence on GPU memory in future studies and even find an approximate solution to realize the algorithm without GPU \cite{GF-CF}.

Third, RevGNN's inference process goes beyond the requirements of the online service. 
If RevGNN is deployed as a real-world application, we need to build appropriate offline and online policies. 
For example, we set the two-stage encoding process as an offline task in the server to periodically produce embedding representations of scholars and papers, and then use the decoder as an online service to recommend potential reviewers for submissions.
Even so, users still have to wait a long time during the cold startup phase, which is another issue to be addressed by future studies.

Fourth, like most existing deep learning models, RevGNN only adapts to the current dataset, meaning a model that performs well on a specific dataset may not generalize well to another dataset. Recently, the emergence of Large Language Models (LLMs) has introduced new techniques for enhancing zero-shot capabilities. LLMs, leveraging the rich knowledge acquired during the pre-training phase, can make recommendations using only a few prompts without updating parameters. Notably, recent works like GPT4Rec~\cite{li2023gpt4rec} and GNERE~\cite{liu2023first} have addressed the cold-start problem of models, using LLMs to achieve breakthroughs and attain excellent performance. Although these approaches have not yet been validated in the context of reviewer recommendation, LLM-based recommendations to address model generalization capability warrant widespread attention.

We also list the following directions for future studies:
\begin{itemize}
    \item Although the Frontiers dataset has significantly addressed the issues of density and scale in traditional reviewer recommendation datasets, its data may still carry biases introduced by the publisher, such as editorial preferences and biased topic selection. In other words, the current reviewer recommendations still require more diverse scenarios and data to assess the generalization and stability of model performance under various conditions, constraints, topics, publishers, etc. Therefore, future research needs to further validate the model's generalizability and stability in different situations.
    \item The augmentation of data naturally necessitates that models be jointly trained across multiple venues, reviewer pools, and corpora to achieve robust generalization. However, this requirement conflicts with the confidentiality required by the peer review process. In recent years, privacy-preserving representation learning methods such as federated learning enable models to be trained across various private datasets in a distributed manner. Models update jointly based on parameter gradients uploaded by each client, thereby mitigating the risk of information leakage inherent in centralized data collection. Therefore, it is imperative to incorporate federated recommendation models based on graph representation learning into future research studies.
    \item Furthermore, with the increase in new data, continuously retraining recommendation models during actual service is both costly and time-consuming. Moreover, new data often brings about distribution shifts in features, leading to performance degradation. On the other hand, continuing training on existing models may also result in the dilemma of catastrophic forgetting issues. Therefore, addressing the challenge of enabling recommendation models to continuously learn from new data without forgetting past features is a crucial requirement in practical application scenarios.
    \item The emergence of LLMs brings new technologies and challenges to reviewer recommendation models. Although LLMs have been proven effective in numerous studies~\cite{lin2023can}, it is a well-known fact that LLMs cannot directly read graph-structured data. Existing works~\cite{li2023gpt4rec,liu2023first} attempt to bridge the gap between LLMs and graph data by textualizing graph structures, but it remains challenging for LLMs to truly comprehend the complex topological structures in graphs. In other domains, such as visual images~\cite{liu2023visual}, some works have converted images into visual tokens, enabling LLMs to understand images while retaining traditional visual encoders. This approach could be insightful for GNN-based recommendation systems.
\end{itemize}

\section{Conclusion}

In this study, we propose the RevGNN, for the academic reviewer recommendation problem.
The RevGNN model uses a two-stage encoder to learn comprehensive embedding representations of reviewers and submissions through multiple aspects including behavior preference, priori scientific semantic knowledge, and contrastive interactions. 
In particular, we investigate the extreme sparsity in the scholar-submission bipartite graph and propose to use the Pseudo Neg-Label strategy to improve the quality of negative sampling during the graph contrastive learning process.
Experimental results show that our proposed RevGNN has achieved a significant performance improvement.
Based on the experimental results, we further analyze and discuss its effectiveness.

\section*{Acknowledgments}
This work was supported by National Natural Science Foundation of China under Grant 62076035 and 62071049, Beijing Natural Science Foundation Project under Grant 4222018, Shanghai Baiyulan Talent Plan Pujiang Project under Grant 23PJ1413800, China Postdoctoral Science Foundation under Grant 2022M711814, and by the Science and Technology Project of State Grid Hebei Information and Telecommunication Branch under Grant SGHEXT00SJJS2200026.


\bibliographystyle{ACM-Reference-Format}
\bibliography{references}


\end{document}